\documentclass[preprint]{revtex4}

\usepackage{graphicx,psfrag}

\begin{document}

\title{Smile dynamics -- a theory of the implied leverage effect: ERRATUM}

\author{Stefano Ciliberti, Jean-Philippe Bouchaud, Marc Potters}
\affiliation{Science \& Finance, Capital Fund Management, 6 Bd Haussmann, 75009 Paris, France}

\date{\today}

\maketitle

We discovered a very unfortunate mistake in our paper ``Smile dynamics -- a theory of the implied leverage effect'' \cite{Paper}, 
where the predictions for the change of implied volatility for a fixed strike and for a fixed moneyness got mixed up. 
When the return of the underlying is $r$, the theory predicts that the at-the-money (ATM) implied volatility $\Sigma_t({\cal M}=0,T)$ 
{\it for fixed moneyness} should evolve as:
\begin{equation}
\frac{\delta \Sigma(0,T)}{\Sigma(0,T)} \approx \frac {r}{2\Sigma(0,T) T} \int_{0}^T du \ g_L(u) 
\equiv \gamma(T) r,  
\end{equation}
where $g_L$ is the \emph{leverage correlation function} of returns $g_L(t) = \langle r_i r^2_{i+t} \rangle_c /\sigma^{3}$. 
This expression should be compared to the original expression in \cite{Paper}:
\begin{equation}
\frac{\delta \Sigma(0,T)}{\Sigma(0,T)} \approx \frac {r}{2\Sigma(0,T) T^2} \int_{0}^T du \, u \ g_L(u),  
\end{equation}
which instead holds for the implied volatility of a {\it fixed strike option} with a moneyness close to zero. 

The correct comparison with empirical data on the OEX index, large cap, mid cap and small cap stocks, 
is given in Figs. 1-4, where we show the implied leverage coefficient $\gamma(T)$ as a function of maturity. 
 The implied data is obtained by regressing the relative
daily change of ATM implied vols on the corresponding stock or index return, for each maturity. The result is then averaged over all stocks within
a given tranche of market capitalisation. The three curves correspond to (a) the 
correct theoretical prediction computed using the historically determined leverage correlation $g_L(t)$; (b) the original theoretical prediction; 
(c) the ``sticky strike'' procedure. We also show the $\gamma = 0$ line corresponding to ``sticky delta''. 

We see that for the OEX index, the implied volatility overreacts to changes of prices compared to the prediction calibrated on the historical leverage effect, except for
the shortest maturities where the prediction is right on the empirical value. For single stocks, small maturity options tend to underreact, whereas longer maturities tend 
to overreact. As mentionned in \cite{Paper}, the empirical curves for the OEX and for large caps appear to be well fitted by a sticky-strike prediction, but with amplitude of the leverage correlation 
substantially larger than its historical value. This would be compatible with the fact that market makers use a simple sticky strike procedure, but with a smile
that is significantly more skewed than justified by historical data, or else that the a local volatility model is used, since this leads to a factor 2 amplification 
of the sticky-strike prediction \cite{Bergomi}. 

The other results of the paper, in particular the empirical work, is unaffected by the above blunder. We thank V. Vargas and L. De Leo for discussions.

\begin{figure}[]
  \label{fig:gammalc}
  \includegraphics[angle=-90,width= 0.7\textwidth]{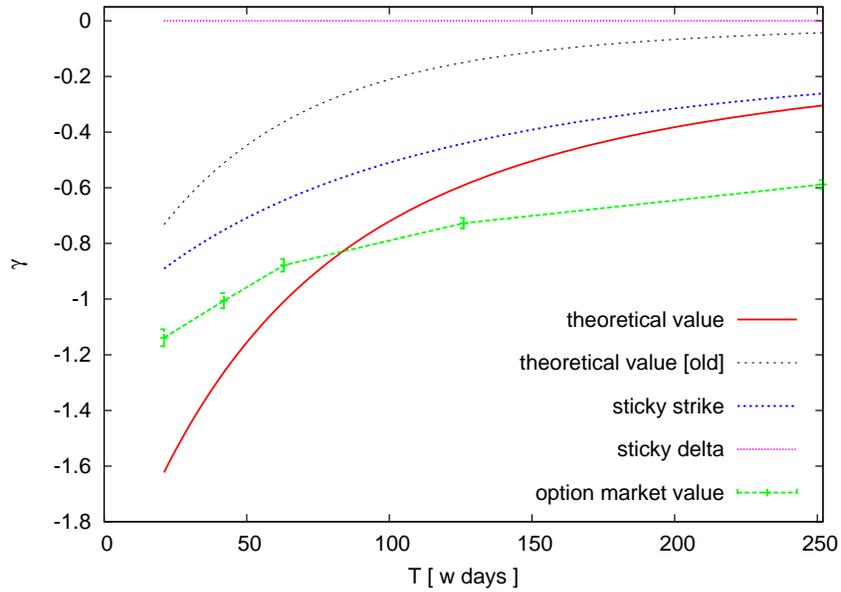}
  \caption{Comparison between the implied leverage coefficient $\gamma$ and the various theoretical predictions, 
  sticky strike, striky delta, historical (wrong) and historical (corrected), for 
  large cap US stocks in the period 2004-2008. 
  }
\end{figure}

\begin{figure}[]
  \label{fig:gammamc}
  \includegraphics[angle=-90,width= 0.7\textwidth]{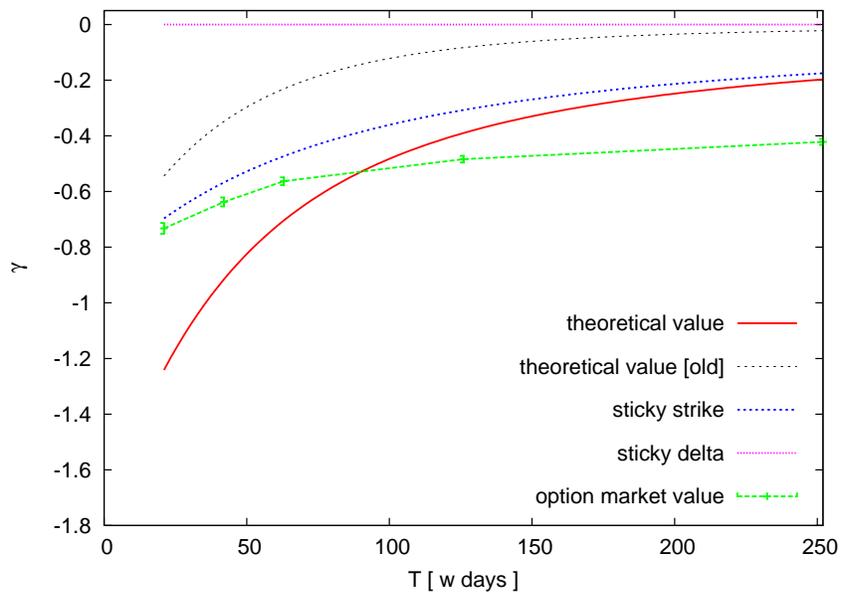}
  \caption{Same as Fig. 1, but for mid-cap US stocks.}
\end{figure}

\begin{figure}[]
  \label{fig:gammasc}
  \includegraphics[angle=-90,width= 0.7\textwidth]{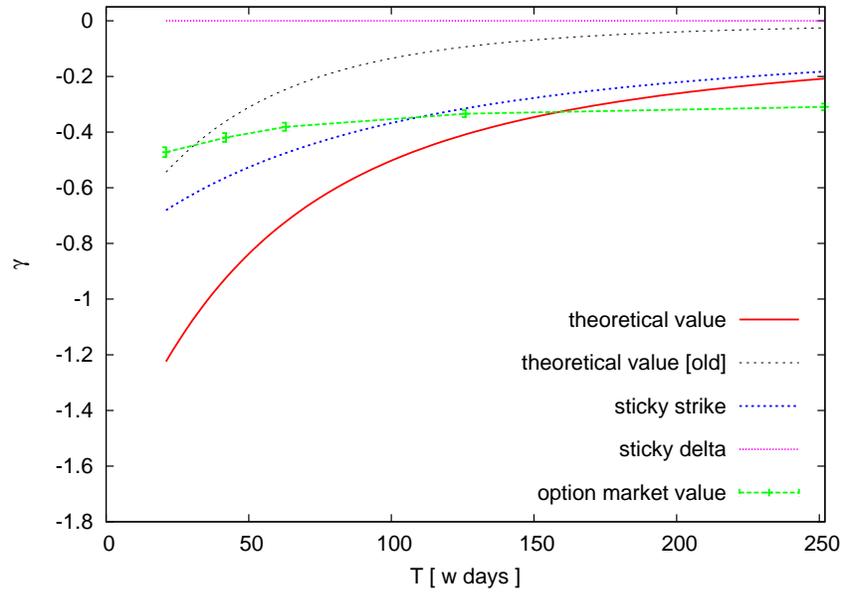}
  \caption{Same as Fig. 1, but for small-cap US stocks. }
\end{figure}

\begin{figure}[]
  \label{fig:oex}
  \includegraphics[angle=-90,width= 0.7\textwidth]{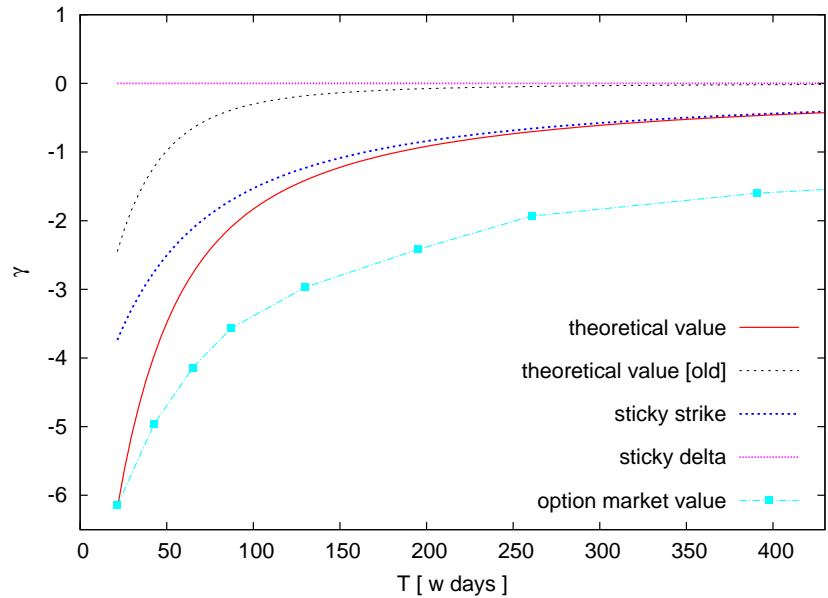}
  \caption{Same as Fig. 1, but for the OEX index. Note that the shortest, most liquid, maturity is perfectly explained by the theory, whereas longer 
  maturities overreact.}
\end{figure}


\begin{thebibliography}{0}

\bibitem{Paper} S. Ciliberti, J.P. Bouchaud, M. Potters, {\it Smile dynamics -- a theory of the implied leverage effect}, Wilmott Journal
Volume 1, Issue 2, pages 87-94, April 2009

\bibitem{Bergomi} see e.g. L. Bergomi, {\it Smile Dynamics IV}, Risk Magazine, December 2009.

\end{thebibliography}
\end{document}